# Transient Effects in Fission Evidenced from New Experimental Signatures[a]

B. Jurado[1,b], C. Schmitt[1], K.-H. Schmidt[1], J. Benlliure[2], T. Enqvist[1,c], A. R. Junghans[3], A. Kelić[1], F. Rejmund[1,b]

[1] *GSI, Planckstraße 1, 64291 Darmstadt, Germany*
[2] *Facultad de Fisica, Univ. de Santiago de Compostela, 15706 S. de Compostela, Spain*
[3] *Forschungszentrum Rossendorf, Postfach 510119, 01314 Dresden, Germany*

A new experimental approach is introduced to investigate the relaxation of the nuclear deformation degrees of freedom. Highly excited fissioning systems with compact shapes and low angular momenta are produced in peripheral relativistic heavy-ion collisions. Both fission fragments are identified in atomic number. Fission cross sections and fission-fragment element distributions are determined as a function of the fissioning element. From the comparison of these new observables with a nuclear-reaction code a value for the transient time is deduced.

**PACS: 24.10.-i, 24.75.+i**

*Introduction.-*The process of equilibration of a highly excited nucleus in all its degrees of freedom is not yet well understood. A dynamical description of the equilibration process in terms of a purely microscopic theory is not possible to the present day due to the large number of degrees of freedom involved. For this reason, most of the current theoretical models are based on transport theories [1], where one distinguishes between collective and intrinsic degrees of freedom. The latter are not considered in detail but in an average sense as a heat bath. The transfer of excitation energy between collective and intrinsic degrees of freedom is denominated dissipation and quantified by the dissipation strength $\beta$, which may depend on excitation energy and deformation.

One of the most intensively investigated nuclear collective motions is the fission process. In the frame of a transport theory, fission is the result of the evolution of the collective fission coordinates under the interaction with the heat bath and an external driving force given by the available phase space. This evolution can be obtained by solving the Langevin equation or its integral form, the Fokker-Planck equation (FPE) [2]. Already in 1940, Kramers [3] described the nuclear fission process within a transport theory and derived the *stationary* solution of the corresponding FPE. Later, by solving numerically the *time-dependent* FPE, Grangé et al. [4] investigated the transient effects that arise from the relaxation of the collective degrees of freedom. Their results showed that it takes a so-called transient time $\tau_{trans}$, until the fission width reaches 90% of its stationary value.

The most frequently applied tools to measure nuclear times are the neutron clock [5] and the gamma clock [6]. They have yielded the majority of the available information on the time a heavy nuclear system needs to cross the scission point. However, the mean scission time is an integral value, including the transient time, the inverse of the stationary decay rate (the statistical decay time) and an additional dynamic saddle-to scission time. Thus it does not give direct access to the transient time that is connected to the equilibration process of the compound nucleus in deformation. On the other hand, total fission or evaporation-residue cross sections have been used to investigate dissipation at low deformation, but, as we will show later, they are not sufficient to determine transient effects in an unambiguous way. Besides, the experimental manifestation of transient effects is subject of controversy nowadays [7]. In fact, the observation of transient effects requires a reaction mechanism that forms excited nuclei with an initial population in deformation space far from equilibrium and an experimental signature that is specifically sensitive to the delayed population of transition states.

---

[a] This work forms part of the PhD thesis of B. Jurado
[b] Present address: GANIL, Blvd. H. Becquerel, B.P. 5027, 14076 Caen, France
[c] Present address: CUPP project, P.O. Box 22, 86801 Pyhäsalmi, Finland

In the present work, highly excited fissioning systems with compact shapes and low angular momenta were produced in peripheral heavy-ion collisions. Both fission fragments were identified in atomic number, enabling the measurement of fission cross sections and fission-fragment element distributions as a function of the fissioning element. This way we introduce two new experimental signatures, which are selectively sensitive to transient effects. They are exploited to deduce a quantitative value for the transient time from a comparison with a nuclear-reaction code, where dissipation effects in fission are modeled in a realistic way. The new approach should help solving the questions on the strength of $\beta$ and its variation with deformation [8, 9, 10, 11, 12] and temperature [9, 13, 14, 15], which are still intensively discussed.

*Experiment.-* In very peripheral nuclear collisions with relativistic $^{238}$U ions from the SIS18 heavy-ion accelerator of GSI, fissile nuclei were produced with high excitation energy and small deformation. This reaction mechanism induces only small angular momenta l < 20$\hbar$ [16], which avoids additional complications in describing the process. The experimental set-up, especially conceived for fission studies in inverse kinematics [17], is schematically illustrated in Fig. 1. In the present work we used a lead target and a $(CH_2)_n$ target. When the projectile fragment fissions, the two fission fragments are emitted in forward direction and detected simultaneously in a double ionization chamber that delivers a very accurate measurement of their nuclear charges. The velocity dependence of the energy-loss signals is corrected by means of the time of flight.

The charge identification of both fission fragments enabled us selecting the fission events according to the excitation energy induced in the nuclear collision. Indeed, the sum $Z_1 + Z_2$ of the nuclear charges of the fission fragments is a very significant quantity, because it is essentially identical to the charge of the fissioning nucleus. Our model calculations described below indicate that the charge of the fissioning system is roughly proportional to the charge of the projectile fragment and hence, it gives a measure of the centrality of the collision. Lower values of $Z_1 + Z_2$ imply smaller impact parameters and higher excitation energies induced by the fragmentation process.

The first signature we exploited to measure transient effects is given by the partial fission cross sections, i.e. the fission cross sections as a function of $Z_1 + Z_2$. At high excitation energies particle decay times become smaller than the transient time and the nucleus can emit particles while fission is suppressed. Therefore, for the lightest fissioning nuclei (lowest values of $Z_1 + Z_2$) transient effects will lead to a considerable reduction of the fission probability. The second signature is based on the charge distribution of the fission fragments that result from a given fissioning element. According to empirical systematics [18], the width of the mass distribution of the fission fragments is a measure of the saddle-point temperature. This experimental result has been corroborated by two-dimensional Langevin calculations [19]. Due to the strong correlation between the mass distribution and the charge distribution of the fission fragments, the same relation holds for the width of the charge distribution and the temperature at saddle. Thus, for the lower values of $Z_1 + Z_2$ where the initial excitation energy is large and fission is suppressed with respect to evaporation, the nucleus will evaporate more particles on its way to fission. Therefore, transient effects will reduce the temperature of the system at saddle, and, consequently, also the width of the corresponding charge distributions [20]. Thus, two nuclear clocks for measuring the delayed population of transition states were established.

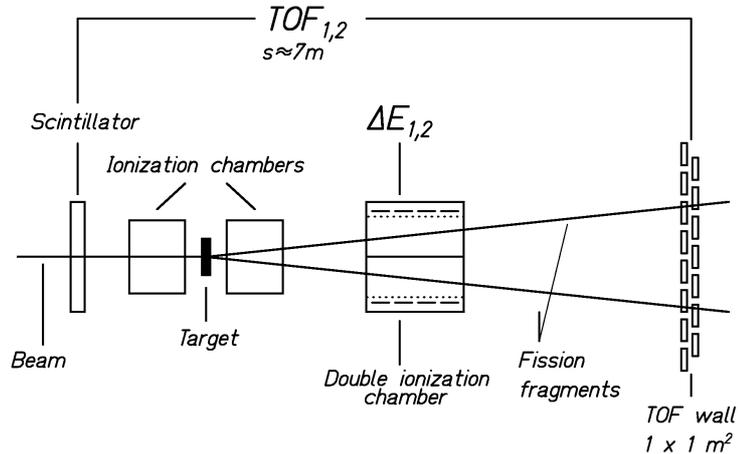

FIG. 1. Experimental set-up for fission studies in inverse kinematics with a 1 $A$ GeV $^{238}$U primary beam.

*Results.-* To deduce quantitative results on transient effects, the experimental observables introduced in the previous section need to be compared with a nuclear-reaction code. The code we use is an extended version of the abrasion-ablation Monte-Carlo code ABRABLA [21, 22]. It describes the nuclear reaction in three stages: In the first stage the characteristics of the projectile residue after the fragmentation are described according to the abrasion model. The second stage accounts for the simultaneous emission of nucleons and clusters (simultaneous break-up) that takes place due to thermal instabilities when the temperature of the projectile spectator exceeds some 5 MeV [23]. After abrasion or eventually the consecutive break-up, the ablation stage models the sequential deexcitation of the system through an evaporation cascade. A reliable study of transient effects requires a dynamical description of the deexcitation process. In particular, the relaxation process in deformation space has to be considered by introducing a time-dependent fission decay-width $\Gamma_f(t)$. For this reason we have implemented in the third stage of ABRABLA a description of $\Gamma_f(t)$ that is based on an approximate solution of the FPE [24]. In Table 1, the experimental total nuclear fission cross section of the reaction of $^{238}$U at 1 $A$ GeV on lead obtained with the set-up shown in Figure 1 is compared with the values obtained from ABRABLA calculations performed with three different treatments of the fission decay width. Apart from the new analytical approximation of [24], Table 1 includes the predictions of two other expressions for the fission decay width, which do not consider any transient effects. In one calculation, we applied the Bohr-Wheeler transition-state model [25], and in the other we used the Kramers solution for the stationary fission-decay width. As expected, the calculation with the transition-state model overestimates the cross section. However, when using the stationary width, the total fission cross section can be reproduced, although with a considerably higher value of the dissipation coefficient than the dynamical calculation given in the last line of

Table 1. These results demonstrate that the total fission cross section allows to identify an overall reduction of the fission probability. However, it does not allow to discriminate between a stationary description of fission and a time-dependent description that includes transient effects.

Table 1. Total nuclear fission cross section of $^{238}$U (1 $A$ GeV) on Pb in comparison with different model calculations.

| Experiment | $\sigma_f^{nucl} = (2.16 \pm 0.14)$ b |
|---|---|
| Bohr-Wheeler | $\sigma_f^{nucl} = 3.33$ b |
| Kramers, $\beta = 6 \times 10^{21}$ s$^{-1}$ | $\sigma_f^{nucl} = 2.19$ b |
| $\Gamma_f(t)$ = FPE [24], $\beta = 2 \times 10^{21}$ s$^{-1}$ | $\sigma_f^{nucl} = 2.09$ b |

Figure 2 represents the partial fission cross sections (Fig. 2a) and the standard deviations of the charge distributions (Fig. 2b) measured in the reaction of $^{238}$U (1 $A$ GeV) on (CH$_2$)$_n$ as a function of the number of protons ($Z_1 + Z_2$) found in the two fission fragments. The experimental data are compared with calculations using the same descriptions for $\Gamma_f(t)$ as in Table 1. In addition, several calculations that include transient effects according to the fission width of [24] with different values of $\beta$ have been performed. As expected, the Bohr-Wheeler transition-state model overestimates both observables, confirming their sensitivity to dissipation. The calculation performed with the constant decay width of Kramers overestimates the observables as well. This means that these new observables are clearly sensitive to transient effects. In particular, the weak increase of the width of the charge distribution with decreasing value of $Z_1 + Z_2$, which goes in line with an increasing excitation energy induced in the nuclear collision, directly proves the suppression of fission at high excitation energies. From the model calculation we deduce that the systems investigated in this work fission only up to excitation energies of about 350 MeV. For both observables, the best description is obtained with $\beta = 2 \cdot 10^{21}$ s$^{-1}$ (full line) corresponding to the critical damping [26] and thus to the shortest transient time $\tau_{trans} \approx (1.7 \pm 0.4) \cdot 10^{-21}$ s.

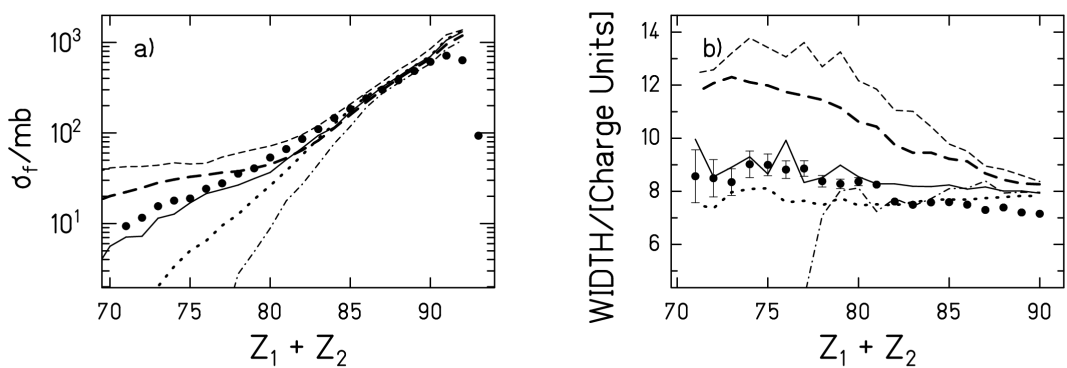

FIG. 2. a) Partial fission cross sections and b) partial widths of the fission-fragment charge distributions for the reaction of $^{238}$U (1 $A$ GeV) on (CH$_2$)$_n$ in comparison with several calculations. The thin dashed line and the thick dashed line are obtained by applying the Bohr-Wheeler transition-state model and the Kramers stationary solution with $\beta = 6 \cdot 10^{21}$ s$^{-1}$, respectively. The full line, the dotted line and the dashed-dotted lines show calculations using the $\Gamma_f(t)$ function of Ref. [24] with $\beta = 2 \cdot 10^{21}$ s$^{-1}$, $\beta = 0.5 \cdot 10^{21}$ s$^{-1}$ and $\beta = 5 \cdot 10^{21}$ s$^{-1}$, respectively. (The staggering in these curves and the strong decrease of the dashed-dotted curve below $Z_1 + Z_2 = 78$ are due to statistical fluctuations of the Monte-Carlo calculations.)

*Conclusion.-* We studied projectile-fragmentation – fission reactions and introduced two new experimental signatures, the partial fission cross sections and the partial widths of the fission-fragment charge distributions, to observe transient effects in fission. These observables exploit the influence of the excitation energy on the fission probability and on fluctuations of the mass-asymmetry degree of freedom. They are based on the particle-emission clock; however, the emission of particles is translated into a reduction of excitation energy before the system passes the fission barrier. We have interpreted the data with a nuclear reaction code that includes a time-dependent treatment of the deexcitation process. The treatment is based on an analytical approximation to the fission-decay width that results from the solution of the Fokker-Planck equation. The comparison of the experimental observables with model calculations indicates that, for the range of temperatures considered, the collective nuclear motion up to the saddle point is critically damped. These new signatures, being sensitive to the dissipation at small deformation, can give new insights into still open questions on the strength of $\beta$ and its variation with deformation and temperature.

In the near future, we plan to extend these investigations to projectiles between uranium and lead in order to separately vary fissility and induced energy (mass, respectively charge loss in the abrasion process) by using secondary beams, presently available at GSI. Further progress in this field is expected when advanced installations, e.g. in the planned GSI or RIA future projects, will become available. They will allow for even more sophisticated fission studies by extending the isospin range of available secondary beams and by adding new capabilities for mass-identification and light-particle-detection, aiming for kinematically complete experiments with a measure of excitation energy in individual events.

*Acknowledgment.-* We thank David Boilley and Anatoly Ignatuk for fruitful discussions. This work has been supported by the European Union in the frame of the HINDAS project under contract FIKW-CT-2000-0031 and by the Spanish MCyT under contract FPA2002-04181-C04-01. One of us (C. S.) is thankful for the financing of a one-year stay at GSI by a Humboldt fellowship. The work profited from a collaboration meeting on "Fission at finite thermal excitations" in April 2002, sponsored by the ECT* ("STATE" contract).